# Structural, magnetic and electronic properties of $CaBaCo_{4-x}M_xO_7$ (M= Fe, Zn)

*V. Cuartero*[*†], J. Blasco*[*‡], G. Subías*[‡], J. García*[‡], J. A. Rodríguez-Velamazán*[‖], C. Ritter*[‖].

[†]ESRF-The European Synchrotron, 71 Avenue des Martyrs, Grenoble (France)

[‡]Instituto de Ciencia de Materiales de Aragón, Departamento de Física de la Materia Condensada, CSIC-Universidad de Zaragoza, C/ Pedro Cerbuna 12, 50009 Zaragoza (Spain)

[‖]Institut Laue-Langevin, Boîte Postale 156, 38042 Grenoble (France)

**Abstract.**

The effect of substituting iron and zinc for cobalt in $CaBaCo_4O_7$ has been investigated using neutron diffraction and x-ray absorption spectroscopy techniques. The orthorhombic distortion present in the parent compound $CaBaCo_4O_7$ decreases with increasing the content of either Fe or Zn. The samples $CaBaCo_3ZnO_7$ and $CaBaCo_{4-x}Fe_xO_7$ with $x \geq 1.5$ are metrically hexagonal but much better refinements in the neutron diffraction patterns are obtained using an orthorhombic unit cell. The two types of substitution have opposite effects on the structural and magnetic properties. Fe atoms preferentially occupy the sites at the triangular layer. Thus, the replacement of Co by Fe supresses the ferrimagnetic ordering of the parent compound and $CaBaCo_{4-x}Fe_xO_7$ ($0.5 \leq x \leq 2$) samples are antiferromagnetically ordered following a new propagation vector

k=(1/3,0,0). However, the Zn atoms prefer occupying the Kagome layer, which is very detrimental for the long range magnetic interactions giving rise to a magnetic glass-like behaviour in the $CaBaCo_3ZnO_7$ sample. The oxidation state of iron and zinc is found to be 3+ and 2+, respectively, independently of the content, as confirmed by x-ray absorption spectroscopy. Therefore, the average Co oxidation state changes accordingly with the $Fe^{3+}$ or $Zn^{2+}$ doping. Also, x-ray absorption spectroscopy data confirms the different preferential occupation for both Fe and Zn cations. The combined information obtained by neutron diffraction and x-ray absorption spectroscopy indicates that cobalt atoms can be either in a fluctuating $Co^{2+}/Co^{3+}$ valence state or, alternatively, $Co^{2+}$ and $Co^{3+}$ ions being randomly distributed in the lattice. These results explain the occurrence of local disorder in the $CoO_4$ tetrahedra obtained by EXAFS. An anomaly in the lattice parameters and an increase in the local disorder is observed only at the ferrimagnetic transition for $CaBaCo_4O_7$ revealing the occurrence of local magneto-elastic coupling.

**INTRODUCTION**

The promise for possible applications in spintronics has motivated and enhanced in the last decade the quest and the understanding of materials showing coupled magnetic and electric properties (1). The presence of polar structures and magnetic frustration seem to be the basic ingredients for the appearance of magnetoelecric coupling. In this respect, the so-called “114” $RBaCo_4O_7$ (R = Y, Ln) compounds, belonging to the family of swedenborgites with the formula $ABM_4O_7$ (M being the transition metal), have been the subject of extensive studies. They show non-centrosymmetric structures with a large variety of magnetic correlations (2–6), best illustrated by the recent emergence of the $CaBaCo_4O_7$ compound (7). The former compounds show generally antiferromagnetic correlations, but more interestingly $CaBaCo_4O_7$ is ferrimagnetic below $T_c$=64K (7) and also shows a linear magnetoelectric coupling below this temperature (8). *Ab-initio* calculations proved that the material is pyroelectric, and the large pyroelectric currents observed were ascribed to exchange-striction effects (9). It adopts an orthorhombic unit cell (space group *Pbn*$2_1$). The crystallographic structure of $CaBaCo_4O_7$ consists of a stacking of alternating triangular (T) and kagomé (K) layers of $CoO_4$ tetrahedra along the **c** axis. There are four different crystallographic sites for Co: Co1 stays at the T layer, while Co2, Co3, and Co4 are in the K layers (7). Figure 1 shows a picture of the crystal structure. The stoichiometric formula corresponds to $CaBaCo_2^{2+}Co_2^{3+}O_7$ and therefore the nominal oxidation state for Co is 2.5+. V. Caignaert *et al.* (7) have proposed a charge ordering of Co atoms in such a way that $Co^{2+}$ occupies the Co2 and Co3 sites whereas a mixed valence $Co^{3+}/Co^{2+}\underline{L}$ state is at the Co1 and Co4 sites. The complex magnetic structure of $CaBaCo_4O_7$, also shown in Fig. 1, is in agreement with the irreducible representation (Irrep) $\Gamma_4$ (7). It consists of a ferromagnetic coupling between Co2 and Co3 spins on zig-zag chains along the **b** axis while Co4 spins are oriented antiparallel to the previous ones and Co1 spins are almost antiparallel to Co2 and Co3 spins. Ferrimagnetism appears because the magnetic moments at the Co1 and Co4 sites are larger than those at the Co2 and Co3 sites.

The substitution of Co with another transition metal may be useful to increase the transition temperature and magnetic coupling. Previous studies on $Zn^{2+}$ substitutions showed the disappearance of ferrimagnetism in favor of antiferromagnetism (AFM) with even less than 3% of doping (10–12). A logic choice for Co sublattice substitution appears to be the neighbor $Fe^{3+}$ with a $3d^5$ electronic configuration. The isostructural *Pbn*$2_1$ $CaBaFe_4O_7$ exhibits a ferrimagnetic transition at $T_C$~270 K with a higher saturated magnetic moment with respect to $CaBaCo_4O_7$ and magnetoelectric coupling below 80 K under high magnetic fields (13-15). The effects on the structural and macroscopic magnetic properties of low Fe doping on $CaBaCo_4O_7$ have been reported lately (16,17). For very low doping concentrations (x<0.05) the presence of Fe is proposed to impose an AFM coupling with Co ions together with phase separation, differently from $Ga^{3+}$ and $Al^{3+}$ doping, which does not affect the arrangement of Co spins (16). However, a full description of the microscopic arrangement of the structural and magnetic structures of $CaBaCo_{4-x}Fe_xO_7$ for higher Fe doping values is lacking, and only a more detailed study of $CaBaCo_2Fe_2O_7$ single crystal found a magnetic transition at 159 K to an AFM ground state (18). This sample, however, exhibited hexagonal symmetry and intrinsic structural disorder.

Hereby a systematic study of the structural, electronic and magnetic properties of $CaBaCo_{4-x}Fe_xO_7$ (x=0.5, 1, 1.5, 2) and $CaBaCo_{4-x-y}Fe_xZn_yO_7$ (x=0, 1, 2 and y=1) series is presented. The effects of Fe and Zn doping on the structural and magnetic long range ordering are investigated by high-resolution powder neutron diffraction (HPND). Besides that, x-ray absorption spectroscopy (XAS) measurements at the Co and Fe K edges have been performed in order to shed light on the evolution of the Fe, Co and Zn electronic state and local structure with doping. The evolution of the structural and magnetic properties with temperature has been also studied by HPND and XAS in selected samples.

**EXPERIMENTAL SECTION**

The $CaBaCo_{4-x}Fe_xO_7$ (x=0, 0.5, 1, 1.5 and 2) and $CaBaCo_{4-x-y}Fe_xZn_yO_7$ (x= 1, 2 and y=1) samples were prepared by ceramic procedures, similarly to the procedure described on reference (7). Stoichiometric amounts of $Co_3O_4$, $Fe_2O_3$ (ZnO), $BaCO_3$ and $CaCO_3$ were mixed and fired at 960ºC overnight. Then, the powders were pressed into pellets and sintered at 1100ºC for 18 *h* and quenched into air. Two samples, $CaBaCo_{2.5}Fe_{1.5}O_7$ and $CaBaCo_2Fe_2O_7$, were also prepared with an annealing in an Ar flow at 1100ºC to test the effect of the atmosphere in the physical properties. In the same way, selected samples were prepared in oxygen ($CaBaCo_3ZnO_7$) or oxygen ($CaBaCo_2Fe_2O_7$ or $CaBaCo_3FeO_7$) current flows. No significant differences were detected.

X-ray diffraction patterns collected at room temperature agreed with the formation of single phase samples. Neutron diffraction experiments were carried out at the Institute Laue Langevin (Grenoble, France) high-flux reactor. The high-resolution powder diffractometer D2B ($\lambda$=1.596 Å) with an angular range $5° \leq 2\theta \leq 160°$ was used to perform crystallographic studies at selected temperatures between 1.5 and 400 K. The high flux powder diffractometer D1B ($\lambda$=2.52 Å) with an angular range $5° \leq 2\theta \leq 128°$ was used to collect thermodiffractograms between room temperature and 2 K for the $CaBaCo_4O_7$ and $CaBaCo_3FeO_7$ samples. The refinements of crystal and magnetic structures were made using the Fullprof program (19). These refinements were performed in two ways, either using the traditional refinement of fractional coordinates or by the refinement of mode amplitudes obtained from the symmetry mode analysis (20, 21). The symmetry mode analysis was made using the ISODISTORT tool (22).

XAS measurements at the Co, Fe and Zn K edges were carried out with a Si (111) double crystal monochromator at the BM23 beamline (23) of the ESRF (Grenoble, France). X-ray absorption near edge structure (XANES) spectra were normalized to unity edge jump using the Athena software from the Demeter package (24). The extraction of the extended x-ray absorption fine structure (EXAFS) $\chi$(k) signals was also performed using the Athena program and the Fourier Transform (FT) curves of the k weighted $\chi$ (k) signals were obtained for the k range [1.8, 12.5], $\Delta k = 0.3$ Å$^{-1}$, using a sinus window. The EXAFS structural analysis was performed using theoretical phases and amplitudes calculated by the FEFF-6 code (25) and fits to the experimental data were carried out in R-space with the Artemis program of the Demeter package (24). The reference compounds for $Co^{2+}$ and $Co^{3+}$ on a tetrahedral oxygen environment are $CaBaCoZnFe_2O_7$ and $YBaCoZn_3O_7$, respectively.

## RESULTS AND DISCUSSION

### A. Crystal structure at room temperature

Room temperature HPND patterns were collected for $CaBaCo_{4-x}Fe_xO_7$ (x= 0.5, 1, 1.5 and 2) and $CaBaCo_3ZnO_7$. The patterns agree with a swedenborgite single phase isostructural to the orthorhombic $CaBaCo_4O_7$ (7). The latter was also measured using the D1B instrument for the sake of comparison. The refined lattice parameters are shown in Fig. 2. The effect of replacing Co by either Fe or Zn induces different changes in the crystal structure. As Co is substituted, the a- and c-axes expand for both kinds of substitutions. However the b-axis contracts in the case of Zn-substitution while it slightly increases for the Fe-based compounds. Overall, the structural effects of Zn-substitution are more important in the ab-plane while the Fe-doping leads to a higher expansion of the c-axis. In both types of substitution, the unit cell volume increases with decreasing the Co-content in the chemical formula. Moreover, the evolution of the c axis and the volume with doping is similar in both series, the last being ascribed to the expansion of c axis. Preliminary structural characterization (26) reported that $CaBaCo_{4-x}M_xO_7$ (M=Fe or Zn; x= 1, 2) adopt the hexagonal structure of the swedenborgite (space group $P6_3mc$). This suggests composition-driven structural transitions in the $CaBaCo_{4-x}M_xO_7$ system. Using x-ray powder diffraction, $CaBaCo_{4-x}Zn_xO_7$ has been reported to possess an orthorhombic cell for x<0.8 and a hexagonal one for x≥0.8 (11) while in the case of the $CaBaCo_{4-x}Fe_xO_7$ series, the transition occurs for x-values between 0.5 and 1 (17). Our results indicate that the orthorhombic distortion decreases with increasing the x-values in both series as can be seen in Fig. 2(a). However, the orthorhombic distortion is clearly still noticeable in the HPND patterns of $CaBaCo_3FeO_7$ contrary to previous results (17). The $CaBaCo_3ZnO_7$, $CaBaCo_{2.5}Fe_{1.5}O_7$ and $CaBaCo_2Fe_2O_7$ samples are metrically hexagonal within the experimental error but the structural refinements using the space group *P6₃mc* yield unsatisfactory results with very poor reliability factors and several diffraction peaks whose intensity cannot be accounted for. The use of the orthorhombic model, *Pbn2₁*, amended these problems so we decided to perform the whole structural analysis using this model.

Therefore, one can obtain reasonable fits of the HPND patterns by using the orthorhombic model and refining fractional coordinates. However, a large statistical error is observed in the z-coordinate of all atoms. This may be related to the fact that we are working with a polar structure and all atoms are in general positions, which increases the error on its determination. Previous studies on $CaBaCo_4O_7$ have overcome this problem by fixing some z-coordinates (7) whose physical meaning could be questionable. For another sample, $CaBaCo_2Fe_2O_7$, the split of some oxygen positions has been proposed (18). In order to gain more insight into the structural details of these compounds, we have made use of a symmetry analysis. Under this approach, one has to refine the amplitudes of certain symmetry modes that relate a distorted structure to the ideal parent structure. We have taken as parent structure of a swedenborgite the non-centrosymmetric hexagonal one with space group *P6₃mc*. Using the ISODISTORT tool, we have explored the distortions able to transform the hexagonal structure into the orthorhombic one. The orthorhombic distortion is related to the hexagonal parent structure with the following lattice vectors: $\mathbf{a_o}$= (0,1,0), $\mathbf{b_o}$=(2,1,0) and $\mathbf{c_o}$=(0,0,-1) with the same origin. The active point for this transition in the first Brillouin zone is **k**=(1/2,0,0). Therefore, the relationships between orthorhombic and hexagonal lattice parameters are: $a_o=a_h$, $b_o=\sqrt{3}b_h$ and $c_o=c_h$. There are active modes which belong to three different Irreps in the hypothetical $P6_3mc \rightarrow Pbn2_1$ transition. The active Irreps are GM1 and GM5 (**k**=0) with 10 and 13 active modes, respectively, and the Irrep M2, **k**=(1/2,0,0), with 15 modes, which must be the primary modes of this transition. The quality of the fits after refining either fractional coordinates or mode amplitudes is similar. However, the second method allowed us to see the origin of the problem: the error in the z-coordinate is still high and it is due to some modes corresponding to the Irrep GM1. One possibility is to nullify the contribution of these modes (similar to fixing fractional coordinates) but we have noticed that the problematic modes of Irrep GM1 show a high correlation among them, which can be removed by adding a simple constrain. In this way, the same parameter is used to refine both the GM1 mode acting on Co at (2a) Wyckoff position in the hexagonal cell (T-layer), and the GM modes acting on Co at (6c) position (K-layer). With this simple link (reducing a free parameter), the statistical error is strongly reduced. Figure 3 shows an example of the refinements obtained by following this procedure and the refined structural parameters can be found in the supplementary information (table S1).

The cation distribution on the T- and K-layers is a critical point to describe the magnetic behavior of the samples. In order to address this problem, the occupancies of the Co(M) atoms have been refined in both layers, keeping constant the nominal composition (i.e. correlating occupancies of different crystallographic sites). The results on M distributions (supplementary information, Table 1) are displayed in the inset of Fig. 3 for the $CaBaCo_{4-x}Fe_xO_7$ series. Our refinements reveal that the substitution of Co with Fe is not completely random. Fe tends to occupy preferentially the sites at the T-layer and then, the concentration of Co atoms decreases more quickly in the T-layer with increasing x in the $CaBaCo_{4-x}Fe_xO_7$ samples. The opposite happens in the $CaBaCo_3ZnO_7$ sample. Here, the refinements reveal that all Zn atoms occupy the K-layer (33% in concentration) and the T-layer is only composed of Co atoms. The preferential occupations of Fe and Zn atoms can be explained on the basis of the steric effects in the swedenborgite cell.

According to our structural refinements (see supplementary information, Table S2), the average distance in the Co(M)$O_4$ tetrahedron increases with the Fe/Zn content. For the iron series, since iron atoms enter as $Fe^{3+}$, this increase can be explained by the decrease of the Co valence state that implies an increase of the Co-O interatomic distances. The difference among the four crystallographic sites is very small. The Co(M)1$O_4$ tetrahedron shows the shortest average Co(M)-O distance independently of the composition whereas the Co(M)4$O_4$ tetrahedron has also a short distance but it increases with the Fe/Zn doping. On the other hand, Co(M)2$O_4$ and Co(M)3$O_4$ tetrahedra show larger average Co(M)-O distances, which remain nearly independent of the doping. This separation in two groups justifies the charge ordering proposed by V. Caignaert *et al.* (7) for the undoped sample. However, this interpretattion must be taken with caution due to the high dispersion of the Co(M)-O distances on each crystallographic site. In any case, our results suggest that Fe is mainly incorporated as $Fe^{3+}$ in the compressed Co(M)1-$O_4$ tetrahedral sites (T-layer) in the $CaBaCo_{4-x}Fe_xO_7$ series whereas Zn occupies indistinctly the three sites of the Kagome layer.

**B. Magnetic ordering at low temperature**

For the sake of comparison between the two magnetic arrangements found in this study, neutron thermodiffractograms were acquired for $CaBaCo_4O_7$ and $CoBaCo_3FeO_7$ at the D1B instrument ($\lambda$=2.52Å) between 2 and 350 K. The neutron thermodiffractograms can be consulted in the supplementary information. Clearly, the magnetic contribution at low temperature is very different for the two samples and this fact leads to different temperature dependences of the unit cell for each sample. Fig. 4 shows the evolution of the refined lattice parameters with temperature for both samples. This evolution is compared with the emergency of long range magnetic ordering. In the case of $CaBaCo_4O_7$, the new magnetic peaks appear around $T_C$=70 K and they agree with a **k**=0 propagation vector. A collinear ferrimagnetic ordering of Co moments with a main component along b-axis (see Fig. 1) accounts for the experimental pattern in agreement with previous studies (7). The occurrence of the ferrimagnetic peaks at $T_C$ is correlated with significant changes in the lattice parameters. Overall, the a- and c-axes decrease with decreasing temperature above $T_C$. However, the b-axis exhibits an unusual expansion on cooling. At $T_C$, the contraction of the a-axis is suddenly interrupted and it remains almost constant below this temperature. In the case of the c-axis, a noticeable peak marks the onset of the magnetic transition. Finally, the b-axis shows an inflection point at $T_C$ and then it remains constant on cooling. The magnetic structure was refined in the frame of the magnetic (Shubnikov's) group *Pb'n2₁'*. Figure 5 shows the temperature dependence of the refined magnetic moments for the four non-equivalent Co atoms. Clearly, different moments are observed for the atoms with AFM coupling at low temperature. Thus, the pair composed by Co2 and Co3 has moments close to 2.25 $\mu_B$/at whereas values of 2.7 and 3.0 $\mu_B$/at are obtained for Co4 and Co1, respectively. This difference has been identified as an indicator of a $Co^{2+}/Co^{3+}$ charge ordering in the lattice (7) but the refined values are well below the theoretical moments expected for the high-spin ions in tetrahedral coordination: 4 and 3 $\mu_B$/at for $Co^{3+}$ and $Co^{2+}$ respectively, even considering hybridization effects. This implies either a complicated mixture of Co (different charges or spin configurations) in each crystallographic site or the lack of a full polarization in the magnetic ground state.

In the case of $CaBaCo_3FeO_7$, the magnetic contribution appears at higher temperature, around $T_N$=150 K. The new magnetic peaks indexed as (*h/3*, *k*, *l*) reveal that the magnetic structure follows the **k**= (1/3, 0, 0) propagation vector. This is true for all $CaBaCo_{4-x}Fe_xO_7$ samples studied in this work (0.5≤x≤2) indicating that the replacement of Co by Fe leads to a drastic change in the magnetic interactions, which gives rise to a new microscopic magnetic arrangement. The lattice parameters of $CaBaCo_3FeO_7$ show different temperature dependencies. The three axes are contracted on cooling from room temperature as expected. At the magnetic transition temperature $T_N$, a clear peak and valley are observed in the a- and b-axes, respectively. Below $T_N$, the contraction of the a-axis is newly quenched. Finally, a small jump is observed in the c-axis at $T_N$.

In order to find the new magnetic arrangement of the $CaBaCo_{4-x}Fe_xO_7$ (0.5≤x≤2) series, a symmetry analysis was carried out using ISODISTORT tool. The search was focused on the point **k**=(1/3, 0 ,0) and two magnetic Irreps, mDT1 and mDT2, yielding 6 possible magnetic (or Shubnikov) groups to account for the magnetic ordering. Among them, only the magnetic group Pb'n'$2_1$ refines successfully the HPND patterns. In this case, the magnetic cell has a triple a-axis with respect to the nuclear cell. There are six magnetic active modes in this magnetic cell, but only two of them are necessary to reproduce the neutron pattern, indicating that magnetic moments lie in the ac-plane and that the y-component is negligible. Furthermore, and according to the symmetry conditions of the magnetic group, each Co(Fe) atom split in three different orbits in the magnetic cell and each orbit has a multiplicity 4. For instance, the Co1(Fe1) atoms located at the T-layer split in Co11(Fe11), Co12(Fe12) and Co13(Fe13) with multiplicity 4 (12 magnetic atoms). The two active magnetic modes correlate the moment components as follow: the x-component of the magnetic moment is zero for Co11, and it has opposite signs for Co12 and Co13 ($\delta m_x$ and $-\delta m_x$, respectively). In the case of the z-component, given a value $\delta' m_z$ for Co11, the other two orbits (Co12 and Co13) have $-1/2\delta' m_z$. Thus, and according to the mIrrep of Pb'n'$2_1$ group, only two parameters are needed to correlate the moments of the 12 Co(Fe) atoms located at the T-layer. In summary, only 8 parameters are needed to model the whole magnetic structure with 48 Co(Fe) atoms. Using this model, the neutron patterns of all $CaBaCo_{4-x}Fe_xO_7$ ($0.5 \leq x \leq 2$) samples were successfully refined. Figure 6 shows the results obtained for $CoBaCo_3FeO_7$ and the refined magnetic moments are summarized in the supplementary information. The resulting magnetic structure is displayed in the figure 7. First of all, there is clear antiferromagnetic coupling between the atoms in the T- and K-layers. This interaction probably prevents the magnetic frustration in the K-layer. In the ac-plane, there is a threefold modulation of the moments, while the projection along the bc-plane reveals a hexagonal pattern in the magnetic arrangement along x- and z-directions. Although Fe tends to occupy preferentially the site in the T-layer, the substitution of Co by Fe is quite random. This randomness seems to be responsible for the new threefold periodic order with magnetic moments far below from the theoretical ones.

The effects of Zn doping are rather different on the magnetic ordering of the parent compound. In the case of $CoBaCo_3ZnO_7$ sample, there is no trace of long range magnetic ordering at low temperatures, as shown on Fig. 8, which compares the HPND patterns at room temperature and 1.5 K. At low temperature, no new magnetic peaks are observed in the HPND pattern of $CoBaCo_3ZnO_7$, but the occurrence of diffuse scattering at $2\theta \sim 20°$ indicates a mere short-range magnetic ordering in this compound. Therefore the substitution of $Co^{2+}$ with the non-magnetic $Zn^{2+}$ in the K-layer is detrimental for the magnetic arrangement.

### C. X-ray absorption spectroscopy results

In order to gain insights into the electronic properties of these compounds, XANES spectra of the $CaBaCo_{4-x}Fe_xO_7$ (x=0, 1, 2) series and the $CaBaCo_3ZnO_7$ compound were measured. In addition, spectra of $CaBaCoZnFe_2O_7$, $YBaCoZn_3O_7$ and $CaBaCo_2FeZnO_7$ samples were collected as nominal references for $Co^{2+}$, $Co^{3+}$ and $Co^{+2.5}$, respectively with tetrahedral local symmetry. The XANES spectra at the Co K edge of all the samples are plotted on Figure 9(a). There is an evolution of both, the position of the Co K absorption edge and the pre-edge peak, to higher energies with decreasing the Fe content for the $CaBaCo_{4-x}Fe_xO_7$ series. This agrees with an oxidation of the Co valence state from 2+ up to a formal 2.5+ in $CaBaCo_4O_7$. The white line intensity also decreases as the formal Co oxidation state increases, while the intensity of the pre-edge feature increases in agreement with the increase of empty *d* states with-respect to $Co^{2+}$. The chemical shift (~2.7 eV) between $CaBaCoZnFe_2O_7$ and $YBaCoZn_3O_7$ reference samples agrees with the presence of only $Co^{2+}$ and $Co^{3+}$ in the $CaBaCoZnFe_2O_7$ and $YBaCoZn_3O_7$ samples, respectively (27). The Fe K edge normalized XANES for Fe doped samples are shown on Fig. 9(b) and and there is no energy shift in between them, being $E_0$=7124.5 eV, corresponding to the first maximum on the derivative. This corresponds to a chemical shift of 12.5 eV with respect to the metallic Fe reference foil in agreement with the presence of only $Fe^{3+}$ in all of the studied samples (28). Similarly, the normalized XANES spectra at Zn K edge on Zn doped samples barely change - see Fig. 9(c) - which indicates that $Zn^{2+}$ on a tetrahedral environment is stable as well in all the compositions. These results confirm that Fe and Zn are incorporated as $Fe^{3+}$ and $Zn^{2+}$ in the whole $CaBaCo_{4-x}Fe_xO_7$ series and the $CaBaCo_3ZnO_7$ compound, respectively.

The average Co oxidation state for the $CaBaCo_{4-x}M_xO_7$ (M=Fe or Zn) samples has been then estimated by assuming the linear relationship between the chemical shift and the oxidation state (29) when using the appropriate references for $Co^{2+}$ and $Co^{3+}$. The resulting oxidation states are summarized on Table 1, together with the formal oxidation state (assuming only $Fe^{3+}$ and $Zn^{2+}$ in the nominal composition). All values are in a reasonable agreement with the expected stoichiometry, within the experimental error bar, although a slight oxygen deficiency cannot be discarded in some compositions. A mixed-valence oxidation state is found for the Co atom in all the studied $CaBaCo_{4-x}M_xO_7$ (M=Fe or Zn) samples. There are two possibilities for this mixed-valence Co state: an intermediate valence state or the existence of $Co^{2+}$ and $Co^{3+}$ ionic species, either nearly randomly distributed in the lattice (no clear charge ordering can be justified by crystallography) or each site temporally fluctuating between $Co^{2+}$ and $Co^{3+}$ valence states. To distinguish between them, a weighted sum of the two reference XANES spectra for $Co^{2+}$ and $Co^{3+}$ was performed. The best-fit XANES spectra are shown in Fig. 10. The agreement between the simulated and the experimental spectra is quite good and the experimental quantification agrees with the results in Table 1 within the error bar. The weighted sum spectra reproduce very well the absorption edge slope of the experimental spectra and also the white line suggesting the existence of $Co^{2+}$ and $Co^{3+}$ formal ionic species, either spatial (random distribution in the lattice) or temporally (fluctuating in each site) separated.

The average local structure around Co ions has been investigated by EXAFS on the aforementioned samples. The forward Fourier Transformed signals (FFT) of the $k^2$- weighted EXAFS signals of the parent compound and Fe and Zn doped samples (x=1) are represented in Fig. 11. The first peak on the FFT curve corresponds to the first coordination shell around Co, that is, a tetrahedral $CoO_4$ environment. The most intense peak at the lowest R value corresponds to $YBaCoZn_3O_7$ in agreement with the shortest Co-O distances ($Co^{3+}$ reference) and the lowest disorder in the Co local environment. The second sharpest peak is observed for the $Co^{2+}$ reference compound, $CaBaCoZnFe_2O_7$. In this case, the peak appears at higher R-values as expected for longer $Co^{2+}$-O bond lengths. The first coordination shell peak is less intense for the rest of the samples and appears at an intermediate position between those of the two references. The only exception is $CaBaCo_2Fe_2O_7$ that shows almost the same first oxygen coordination shell as the $CaBaCoZnFe_2O_7$ reference sample, in agreement with the deduced $Co^{2+}$ oxidation state from the XANES analysis (see Table 1 and Fig. 10). The references exhibit very different features in the second coordination shell (between 2.5 and 3.5 Å) that are related to the different sites occupied by $Co^{3+}$ (T-layer) or $Co^{2+}$ (K-layer).

In order to correlate the local disorder of the $CoO_4$ tetrahedra on the different compositions with the type and position of the dopant in the crystal structure, we have analyzed the first oxygen coordination shell around Co considering a regular $CoO_4$ tetrahedron with one unique average distance Co-O and one Debye-Waller (D-W) factor. The results are summarized in Table II. The D-W factor includes both thermal and static disorder contributions. Since the EXAFS data shown in Table 2 correspond to the lowest temperature (35 K), the resulting D-W factors probe the average local disorder of the $CoO_4$ tetrahedron. The largest D-W factors are found for the intermediate $Co^{3+/2+}$ valences indicating either a larger distortion or the occupation of different crystallographic sites (disordered distribution). The lowest D-W factors correspond to $YBaCoZn_3O_7$, where cobalt atom is in the 3+ oxidation state and mostly occupies the tetrahedron at the T-site and $CaBaCoZnFe_2O_7$ with only $Co^{2+}$. This last case is a bit different since the two types of dopant coexist in the sample and their preferential occupations (T- or K-layer) are opposite to each other. Thus, $Co^{2+}$ is distributed over more than one crystallographic site and its local environment is slightly more disordered according to the obtained D-W factor. Taking full advantage of the reference compounds, we have calculated the Backward Fourier Transform (BFT) to filter the EXAFS signal corresponding to the 1st coordination shell. The signals from $CaBaCoZnFe_2O_7$ and $YBaCoZn_3O_7$ have been used to probe the Co valence state. In the former sample, Co is 2+ mainly located in the K-layer while for the latter compound Co is 3+ only located in the T-layer. The best weighted sum compared to the experimental signal of a given sample indicates the Co valence state in such sample. The results can be seen in the supplementary information (Fig. S2) and, overall, the Co valence state agrees with the previous estimation using XANES spectra. In this way, the larger D-W factor for the intermediate doped samples can be interpreted as due to the presence of two Co-O interatomic distances corresponding to $Co^{3+}$ and $Co^{2+}$ tetrahedra. This result confirms then the presence of two Co valences in all the doped $CaBaCo_{4-x}M_xO_7$ (M=Fe, Zn) samples either randomly distributed or temporally fluctuating over the four distinct crystallographic sites in the orthorhombic lattice.

Complementary data have been obtained from the analysis of the first oxygen coordination shell around the Fe atom in the $CaBaCo_{4-x}Fe_xO_7$ (x=1, 2) and $CaBaCo_2Fe_2ZnO_7$ samples and the results are also summarized in Table 2. The same average Fe-O interatomic distance is found in all the three samples for the $FeO_4$ tetrahedron according to the presence of only $Fe^{3+}$ (30). Regarding the D-W factors, the higher value is obtained for $CaBaCo_2Fe_2O_7$, being of the same order as that of the $CoO_4$ tetrahedra. Therefore, the same distribution of Co and Fe atoms over more than one crystallographic site takes place in this sample. For the other two Fe-doped samples, very low Debye-Waller factors are determined for the $FeO_4$ tetrahedra, indicating a regular oxygen environment in contrast with the highly disordered one around the Co atoms. This result indicates that Fe atoms prefer to fully occupy one crystallographic site first, confirming the results from neutron diffraction of a preferential incorporation on the T-layer while Co atoms are distributed among the remaining crystallographic sites.

The temperature dependence of EXAFS spectra of $CaBaCo_4O_7$, $CaBaCo_{4-x}Fe_xO_7$ (x=1, 2) and $CaBaCo_3ZnO_7$ samples has been studied from 35 K up to room temperature (RT) at the Co K-edge. The FFT of the $k^2$-weighted EXAFS spectra at selected temperatures is plotted on Fig. 12(a). The major changes appear on the peaks corresponding to the second coordination shell, involving mainly Co-Ba/Ca and Co-Co/M (M: Fe, Zn) scattering paths. The structural analysis have been performed including only single scattering paths up to 4 Å, i.e. Co-O, Co-Ba/Ca, Co-Co/M and using a model that considers one Co atom with the local environment corresponding to the average of the distinct crystallographic sites in the orthorhombic structure. In addition, following our results from neutron diffraction and first-shell EXAFS analysis, for the Fe-doped samples we consider that Fe fully occupies the T-site and the rest is occupied by Co. Other fits with Fe ions on the other crystallographic sites on the initial model were tested and gave always worse fitting parameters. In the case of Zn doping, it is considered that it is placed on K-layer sites. The corresponding best-fits and experimental data at 35 K are shown on Figure 12(b). The excellent agreement validates the proposed structural model and makes the use of further contributions unnecessary.

Upon heating, the Co-O, Co-Co(M) and Co-Ba/Ca distances are barely changing with temperature within the experimental error bar for all compositions (see supplementary information Fig. S3). Figure 13 shows the thermal dependence of the D-W factors for the Co-O and Co-Co contributions. The very low dependence with temperature indicates that the main contribution to the D-W factors for the Co-O distance comes from the structural disorder. For the Fe and Zn doped samples, it slightly decreases with decreasing the temperature remaining nearly constant at low temperatures. However, in the $CaBaCo_4O_7$ sample, it shows an anomalous increase on cooling below 70 K. Therefore, the higher D-W factor observed below 70 K only on the $CaBaCo_4O_7$ parent compound can be ascribed to a more distorted average local environment around Co ions below the ferrimagnetic transition temperature $T_C$. At the same time, the strong change in electric polarization below $T_C$ can be explained by the average increase of the local $CoO_4$ distortion due to magnetostriction mechanisms. On the other hand, no anomalous behaviour was observed for the D-W factors of the average Co-Co interatomic distance in either the doped samples or the parent compound. They decrease with decreasing the temperature following the evolution expected for standard thermal vibrations. Therefore, the local structure instability associated with the magneto-electric transition in $CaBaCo_4O_7$ mainly concerns the $CoO_4$ tetrahedra.

## CONCLUSIONS

The present structural, electronic and magnetic study of the atomic substitution of cobalt atom by Fe and Zn in the pyroelectric $CaBaCoO_4O_7$ compound has allowed us to provide new inputs to the understanding of this system. The structural data obtained from neutron powder diffraction have shown highly distorted $Co(M)O_4$ tetrahedra. In this way, the standard deviation for individual Co(M)-O bond lengths in one tetrahedron is much bigger than the ones observed for the average distances of the four tetrahedra, and this makes crystallographic Co(M) sites very similar to each other. The average Co(M)-O interatomic distances increase with either the Fe or Zn doping. Fe atoms preferentially substitute the Co atoms at the T-layer (Co1 site) whereas Zn atoms occupy the K-layer (Co2, Co3 and Co4 sites). The reason for these different substitutions is related to the available space as $MO_4$ (M=Co, Fe or Zn) tetrahedra are smaller at the T-layers.

XAS demonstrates that Fe is incorporated as $Fe^{3+}$ whereas Zn enters as $Zn^{2+}$. In agreement with these results, the average Co valence state decreases with the Fe doping and increases with the Zn doping, as confirmed by the XAS at the Co K edge. According to the XANES analysis, the mixed-valence state of Co can be well described as a weighted addition of $Co^{3+}$ and $Co^{2+}$, i.e. two valence states, $Co^{2+}$ and $Co^{3+}$, exist. Due to the fact that neutron diffraction cannot show a clear segregation of two $Co^{2+}$ and $Co^{3+}$ sites, we can discard the occurrence of charge ordering. Instead, either a random distribution of $Co^{2+}$ and $Co^{3+}$ in the lattice or a temporally fluctuating valence state at each Co site takes place. This result has been confirmed by the EXAFS analysis. The Fe-O interatomic distances remain constant independently of doping and the D-W factor for the Co-O interatomic distances is maximum for the undoped and low-doped samples, where the higher mixing of the two valence states occurs. This agrees with the atom occupation obtained from neutron diffraction, the occupation of $Co^{3+}$ at the T-layer and the distances at the Co(M)4 site being lower than at the Co(M)2 and Co(M)3 K-sites for the undoped and low-doped samples. Moreover, the EXAFS spectra of the first coordination shell around the Co atoms can be well reproduced by the weighted addition of the correspondent spectra of the $CaBaCoZnFe_2O_7$ ($Co^{2+}$) and $YBaZn_3CoO_7$ ($Co^{3+}$) reference samples, with $Co^{2+}$ mainly in the K-layer and $Co^{3+}$ in the T-layer, respectively.

This site-selective substitution has strong implications on the magnetic and electronic properties of the doped compounds. $CaBaCo_4O_7$ is a collinear ferrimagnet (k=0) along the *ac*-plane, where ferrimagnetism comes from the different magnetic moments of the Co1, Co4 and Co2,Co3 sites, identifying the different intermediate valence states of the these sites. The partial substitution of $Co^{3+}$ with $Fe^{3+}$ at the T-layers (Co1 site), changes the magnetic interactions and a new magnetic structure appears following the propagation vector **k**=(1/3,0,0). The neutron diffraction results show a modulation of the magnetic moment amplitudes. This magnetic structure occurs for all the Fe-doped samples in the studied composition range. It can be noted that the localized moments are far below from the expected ones for the Co/Fe atoms. We think there is a modulated antiferromagnetic order superimposed to a random distribution of both, Fe/Co atoms and $Co^{3+}/Co^{2+}$ valence states, leading to a partial magnetic polarization in each crystallographic site. The occurrence of this threefold periodicity independently of the Fe doping should be matter of a deeper study. On the other hand, the replacement of magnetic $Co^{2+}$ with diamagnetic $Zn^{2+}$ at the K-layers lead to a weakening of the magnetic interactions and prevents long range magnetic ordering in $CaBaCo_3ZnO_7$.

The evolution of the lattice parameters as a function of temperature shows a discontinuity near the magneto-electric transition temperature for the parent $CaBaCo_4O_7$ and the antiferromagnetic transition temperature for $CaBaCo_3FeO_7$. EXAFS measurements as a function of temperature show the presence of a local disorder highly localized in the $CoO_4$ tetrahedra that remains unchanged for all the samples except for the parent $CaBaCo_4O_7$ compound. A small but appreciable increase in the local distortion of the $CoO_4$ tetrahedra is observed at the magneto-electric transition temperature for this sample. This reveals the occurrence of a local magneto-elastic coupling at the ferrimagnetic phase that may be related to the observation of the pyroelectric effect in this composition.

**AUTHOR INFORMATION**

**Corresponding Authors**

* E-mail: cuartero@esrf.fr

*E-mail: jbc@unizar.es

## ACKNOWLEDGMENTS

The authors would like to acknowledge ILL and D2B and D1B-CRG (MINECO) stations for beam time allocation (experiment code 5-31-2366, doi:10.5291/ILL-DATA.5-31-2366). We also acknowledge the ESRF for granting beam time, BM23 staff for technical assistance and Olivier Mathon and Sakura Pascarelli for their support on the XAS experiments. Authors also acknowledge the use of Servicio General de Apoyo a la Investigación-SAI, Universidad de Zaragoza. For the financial support, we thank the Spanish Ministerio de Economía y Competitividad (MINECO), Project MAT2015-68760-C2-1-P and Diputación General de Aragón (DGA), Project E69.

## TABLES

**Table 1.** Experimental chemical shift and corresponding Co valence state for all the $CaBaCo_{4-x}M_xO_7$ (M=Fe or Zn) and $YBaCoZn_3O_7$ compounds.

| **Sample** | **Co formal valence state** | **Chemical shift (eV)** | **Co valence state** |
|---|---|---|---|
| $CaBaCoZnFe_2O_7$ | 2 | 0.0 | 2 |
| $CaBaCo_2Fe_2O_7$ | 2 | 0.18 | 2.07(5) |
| $CaBaCo_3FeO_7$ | 2.33 | 0.47 | 2.17(5) |
| $CaBaCo_2FeZnO_7$ | 2.5 | 1.05 | 2.39(5) |
| $CaBaCo_4O_7$ | 2.5 | 1.12 | 2.41(5) |
| $CaBaCo_3ZnO_7$ | 2.66 | 1.43 | 2.53(5) |
| $YBaCoZn_3O_7$ | 3 | 2.72 | 3 |

**Table 2.** Results from the EXAFS fits of the first coordination shell for the Co(*Fe*)$O_4$ tetrahedron at T=35 K: average Co(*Fe*)-O interatomic distances and their Debye-Waller factors.[1]

| **Co K-edge** | | | | |
|---|---|---|---|---|
| Sample | R <Co-O> (Å) | $<\sigma^2>_{Co-O}$ ($Å^2$) | $\Delta E_0$ (eV) | R fit |
| $CaBaCoZnFe_2O_7$ | 1.94 (1) | 0.0044 (6) | -2.4 (8) | 0.009 |
| $CaBaCo_2Fe_2O_7$ | 1.93 (1) | 0.0042 (3) | -2.8 (4) | 0.011 |
| $CaBaCo_3FeO_7$ | 1.92 (1) | 0.0069 (4) | -4.0 (8) | 0.010 |
| $CaBaCo_2FeZnO_7$ | 1.89 (1) | 0.0074 (6) | -2.5 (4) | 0.005 |
| $CaBaCo_4O_7$ | 1.90 (1) | 0.0076 (6) | -2.0 (9) | 0.011 |
| $CaBaCo_3ZnO_7$ | 1.88 (1) | 0.0066 (5) | -5.3 (6) | 0.007 |
| $YBaCoZn_3O_7$ | 1.84 (1) | 0.0024 (8) | -3.3 (9) | 0.006 |
| **Fe K-edge** | | | | |
| Sample | R <Fe-O> (Å) | $<\sigma^2>_{Fe-O}$ ($Å^2$) | $\Delta E_0$ (eV) | R fit |
| $CaBaCo_2Fe_2O_7$ | 1.88(1) | 0.004(1) | -2.6(18) | 0.010 |
| $CaBaCo_3FeO_7$ | 1.87(1) | 0.002(1) | -2.6(20) | 0.014 |
| $CaBaCo_2FeZnO_7$ | 1.87(1) | 0.003(1) | -3.6(18) | 0.010 |

[1] $S_0^2$ was fixed to 0.74 as a result from the fit of the $Co^{3+}$ reference compound. The fit was performed in *R* space weighted in *k 1*,2,3, being $R_{min}$ = 1Å , $R_{max}$ = 1.9Å , and *dR* = 0.3 and using a sine window.

**FIGURES**

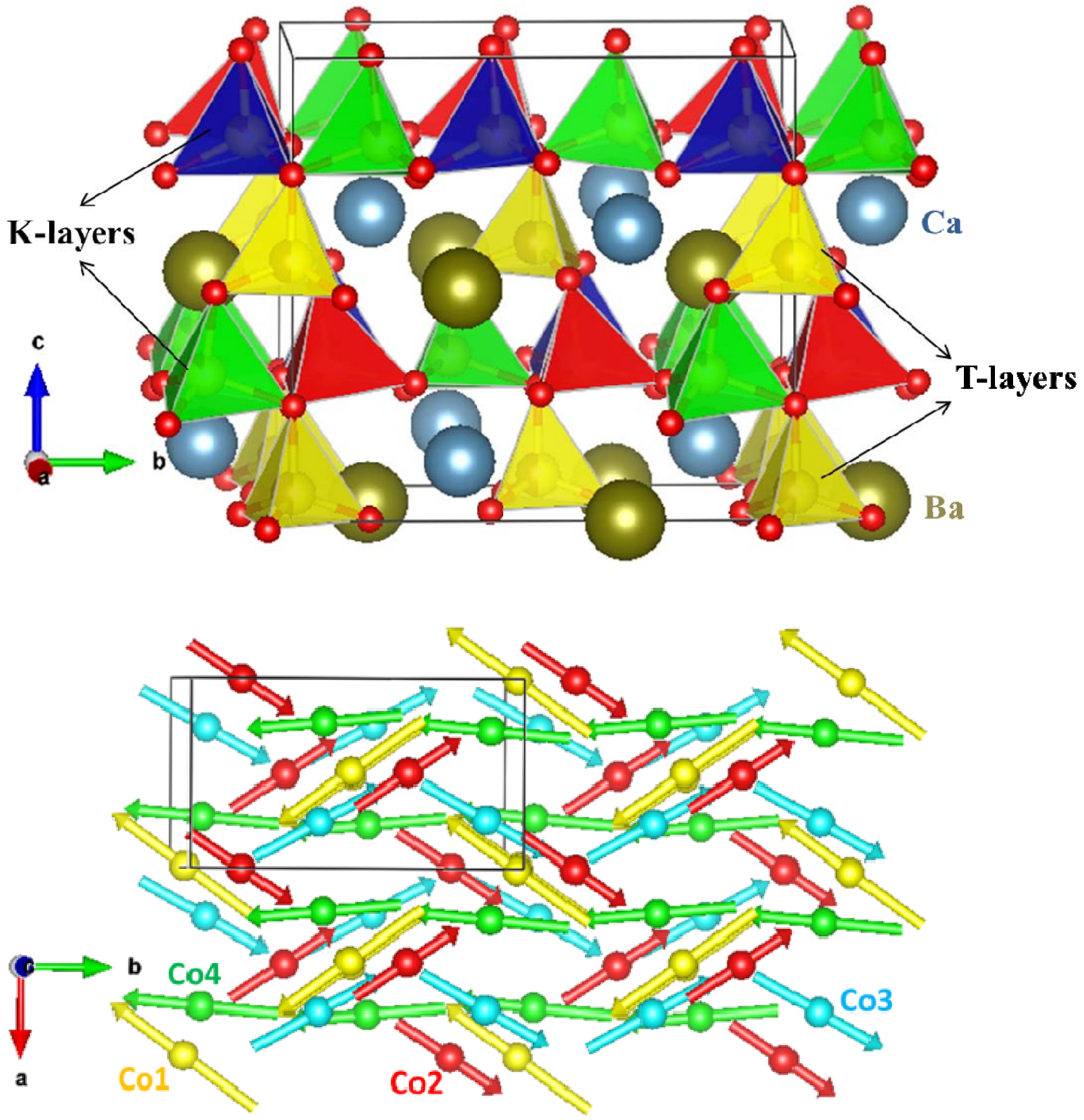

**Figure 1.** Top: Crystal structure of $CaBaCo_4O_7$ showing the stacking of $CoO_4$ tetrahedra forming T- and K-layers. Yellow, red, blue and green tetrahedra stand for Co1, Co2, Co3 and Co4, respectively, in the orthorhombic unit cell. Bottom: Magnetic structure of $CaBaCo_4O_7$ projected onto ab-plane. Yellow, red, blue and green arrows correspond to Co1, Co2, Co3 and Co4 moments, respectively.

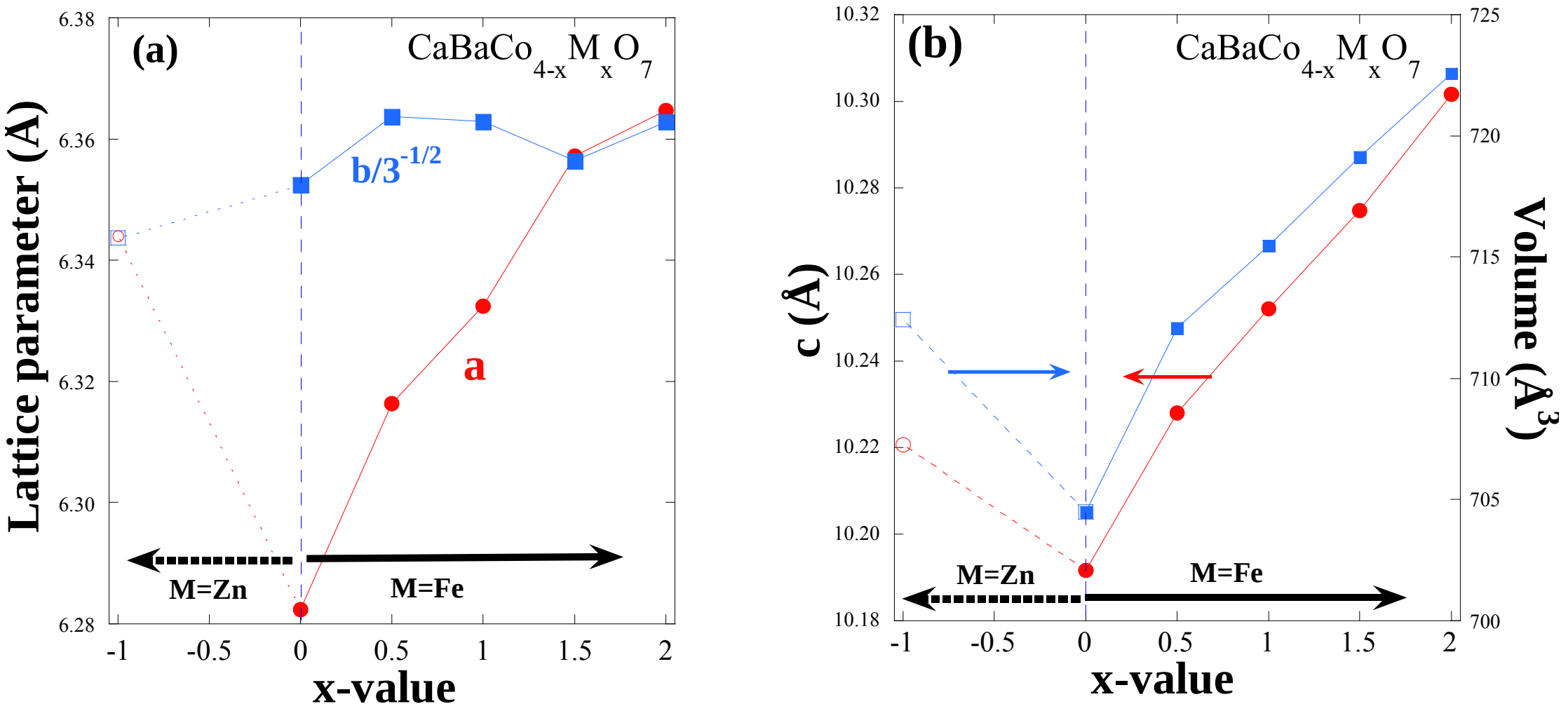

**Figure 2.** Comparison of the lattice parameters for $CaBaCo_{4-x}Fe_xO_7$ and $CoBaCo_{4-x}Zn_xO_7$ series. Panel (a) shows the evolution of a- and b-axes while panel (b) displays c-axis and unit cell volume. Closed and open symbols refer to Fe- and Zn-doped samples, respectively. Data for $CaBaCo_4O_7$ has been obtained from D1B data.

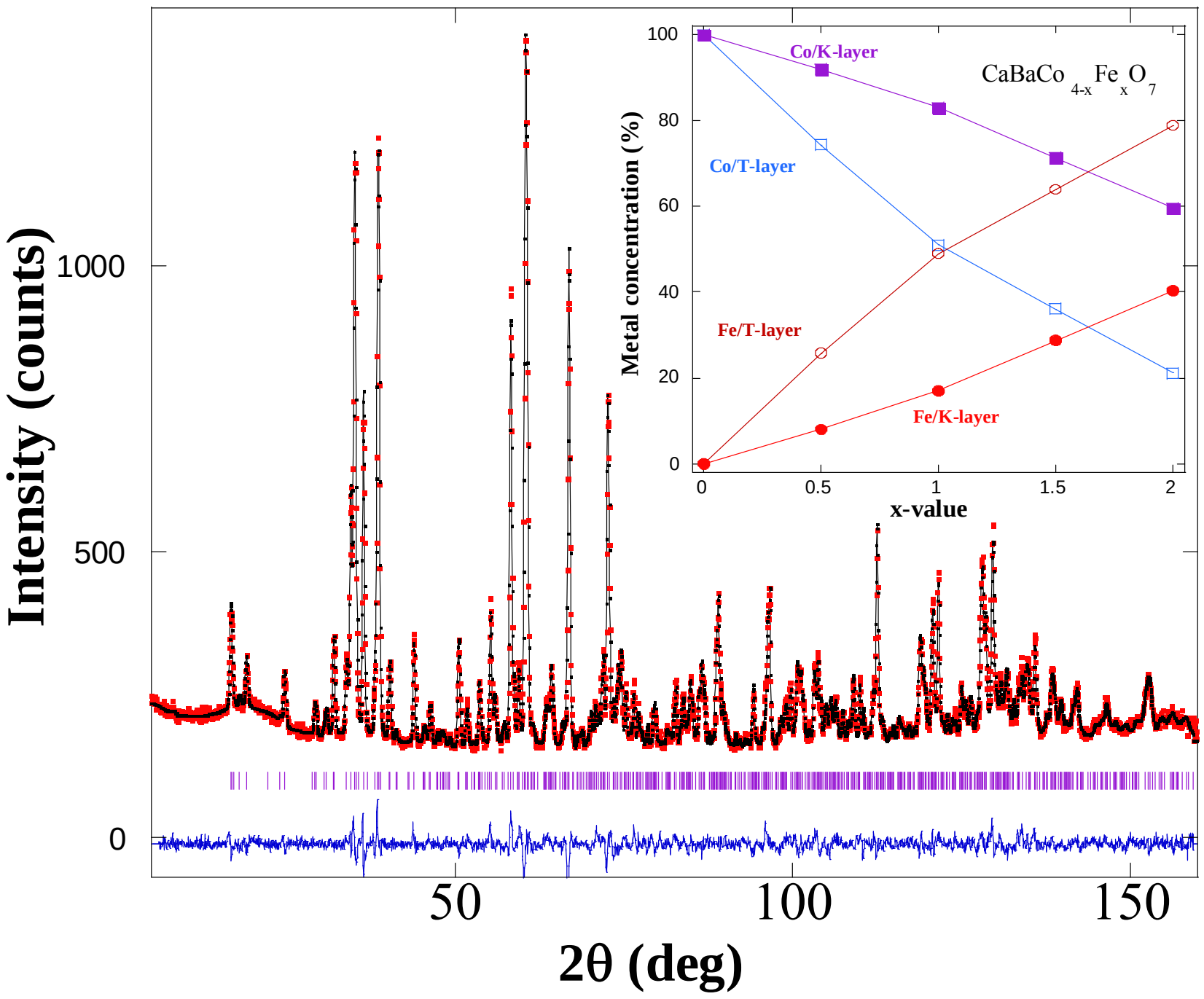

**Figure. 3.** Rietveld refinement of the HPND pattern of $CaBaCo_3FeO_7$ sample at room temperature. Inset: Metal distribution in K- and T-layers for the $CaBaCo_{4-x}Fe_xO_7$ samples.

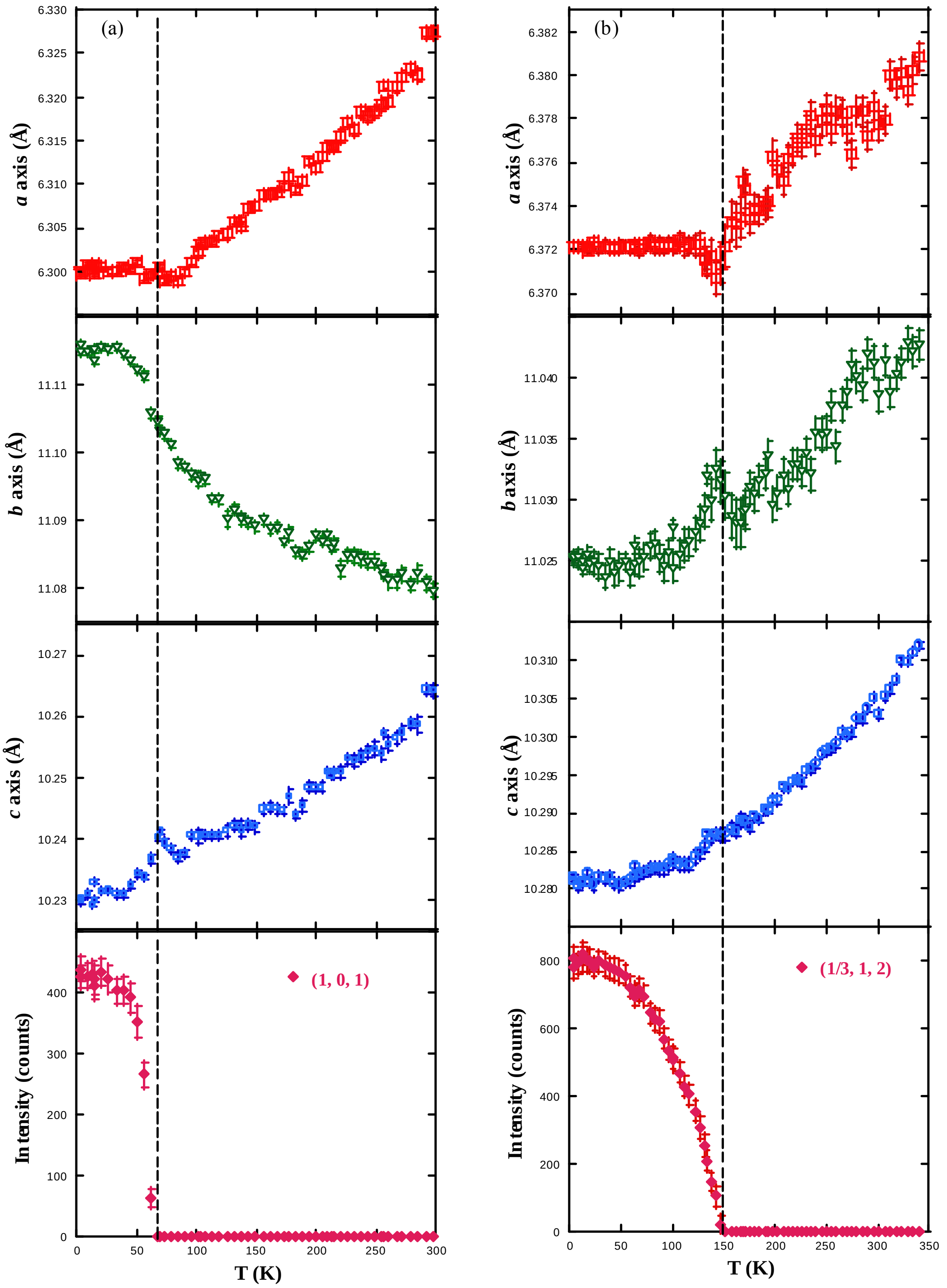

**Figure 4.** Temperature dependence of the refined cell parametes and the integrated intensity of the most intense magnetic peak for (a) $CaBaCo_4O_7$ and (b) $CaBaCo_3FeO_7$.

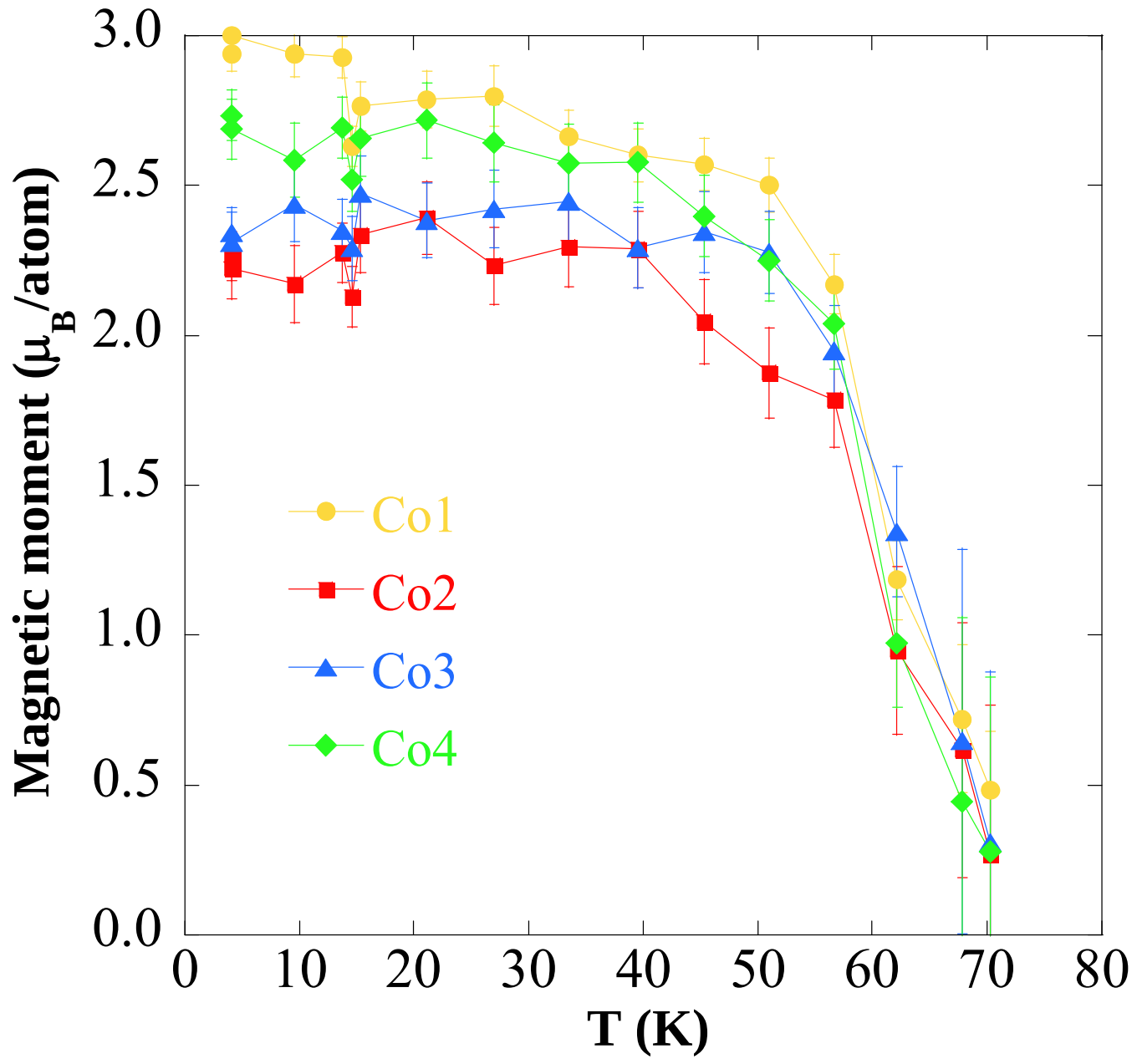

**Figure 5.** Temperature dependence of the refined magnetic moments for the four non-equivalent Co atoms in the $CaBaCo_4O_7$ lattice.

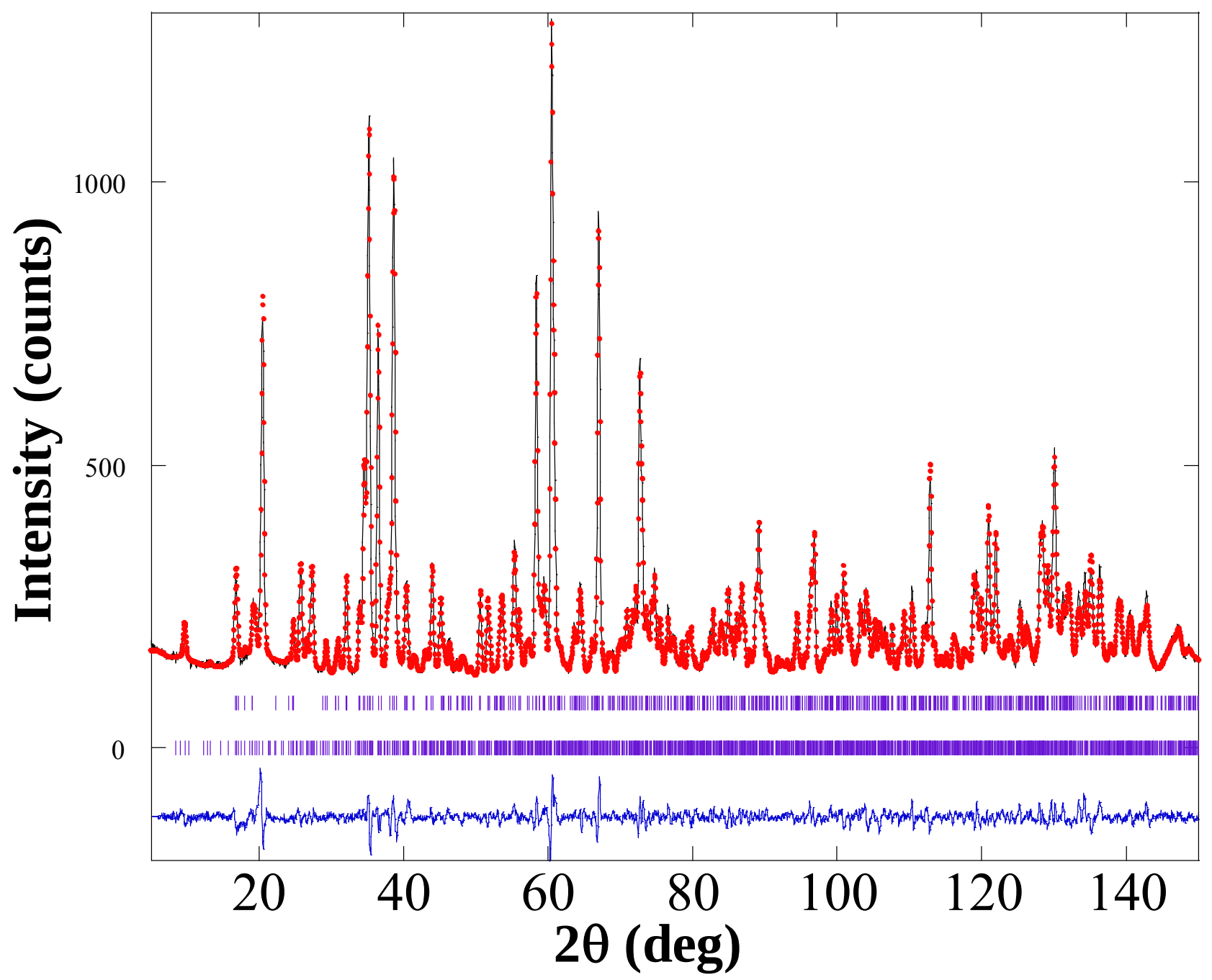

**Figure 6.** Rietveld refinement of the HPND pattern of $CaBaCo_3FeO_7$ sample at 1.5 K

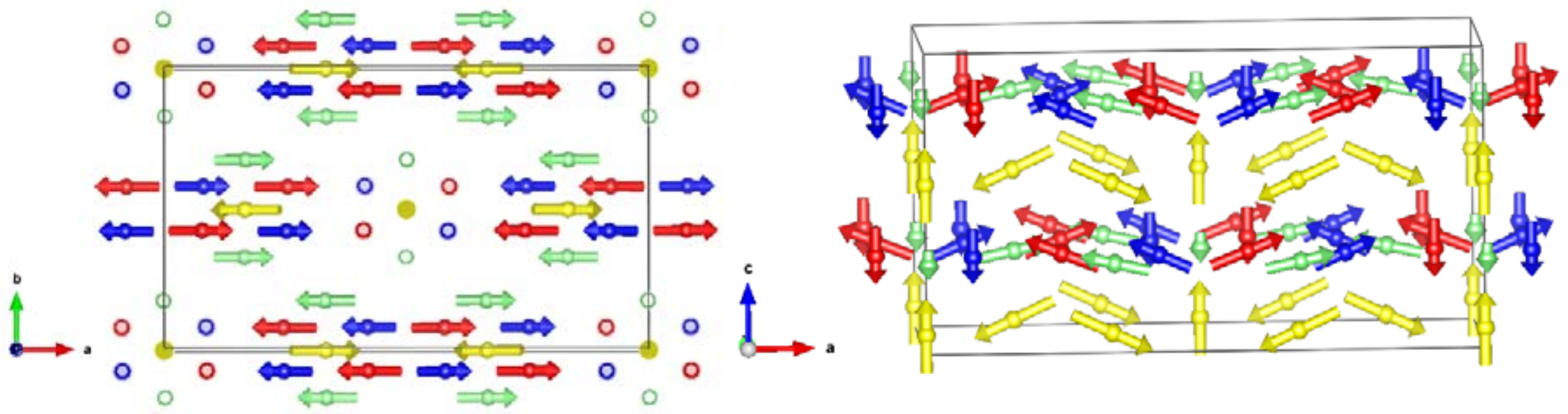

**Figure 7.** Magnetic structure of $CaBaCo_{4-x}Fe_xO_7$ ($0.5 \leq x \leq 1$) projected onto ab-plane (left) and ac-plane (right). Yellow spins corresponds to Co1(Fe1) atoms in the T-layer while the rest are at the K-layers. Red, blue and green moments correspond to Co2, Co3 and Co4 sites, respectively

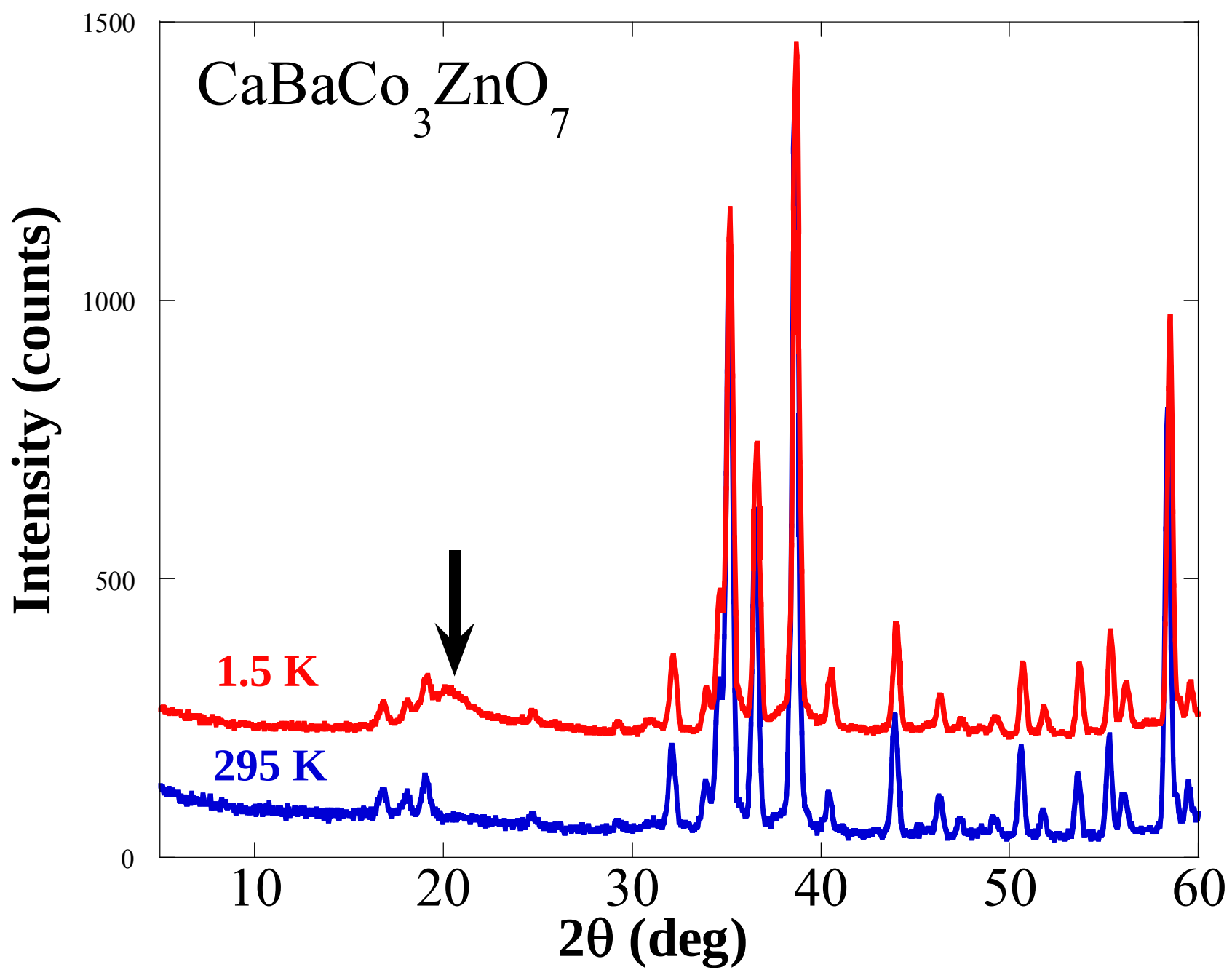

**Figure 8.** Details of the HPND patterns of $CaBaCo_3ZnO_7$ collected at 1.5 and 295 K. The arrow shows the diffuse scattering observed at low temperature indicating short range magnetic ordering

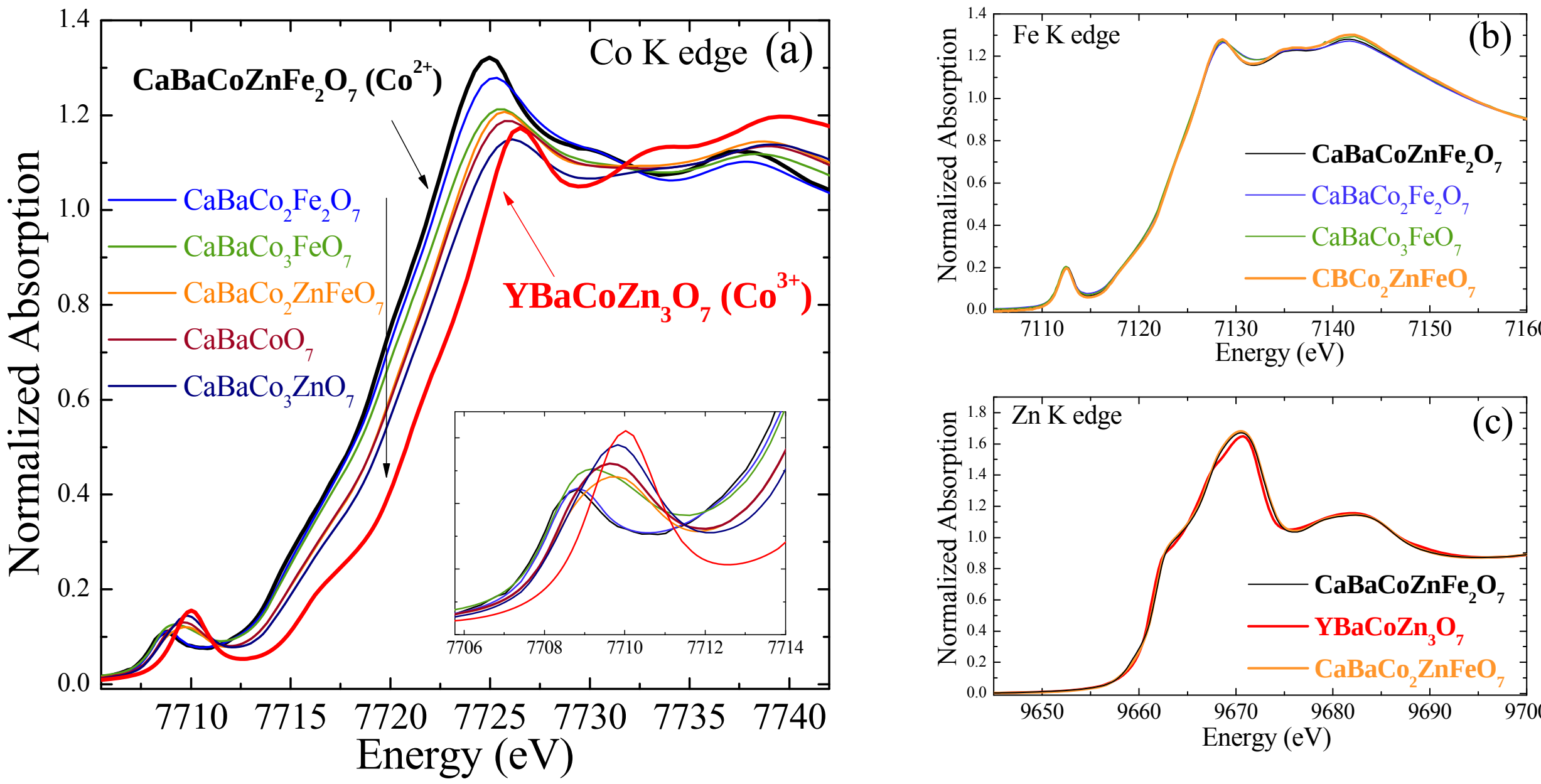

**Figure 9.** Normalized XANES spectra for $CaBaCoO_7$, $CaBaCo_{4-x}Fe_xO_7$ (x=1, 2), $CaBaCo_3ZnO_7$ and $CaBaCo_2ZnFeO_7$ at the Co K-edge **(a)**, including $Co^{2+}$ (black) and $Co^{3+}$ (red) references, and a zoom of the pre-edge region in the inset; at the Fe K-edge **(b)** and at the Zn K-edge **(c)**.

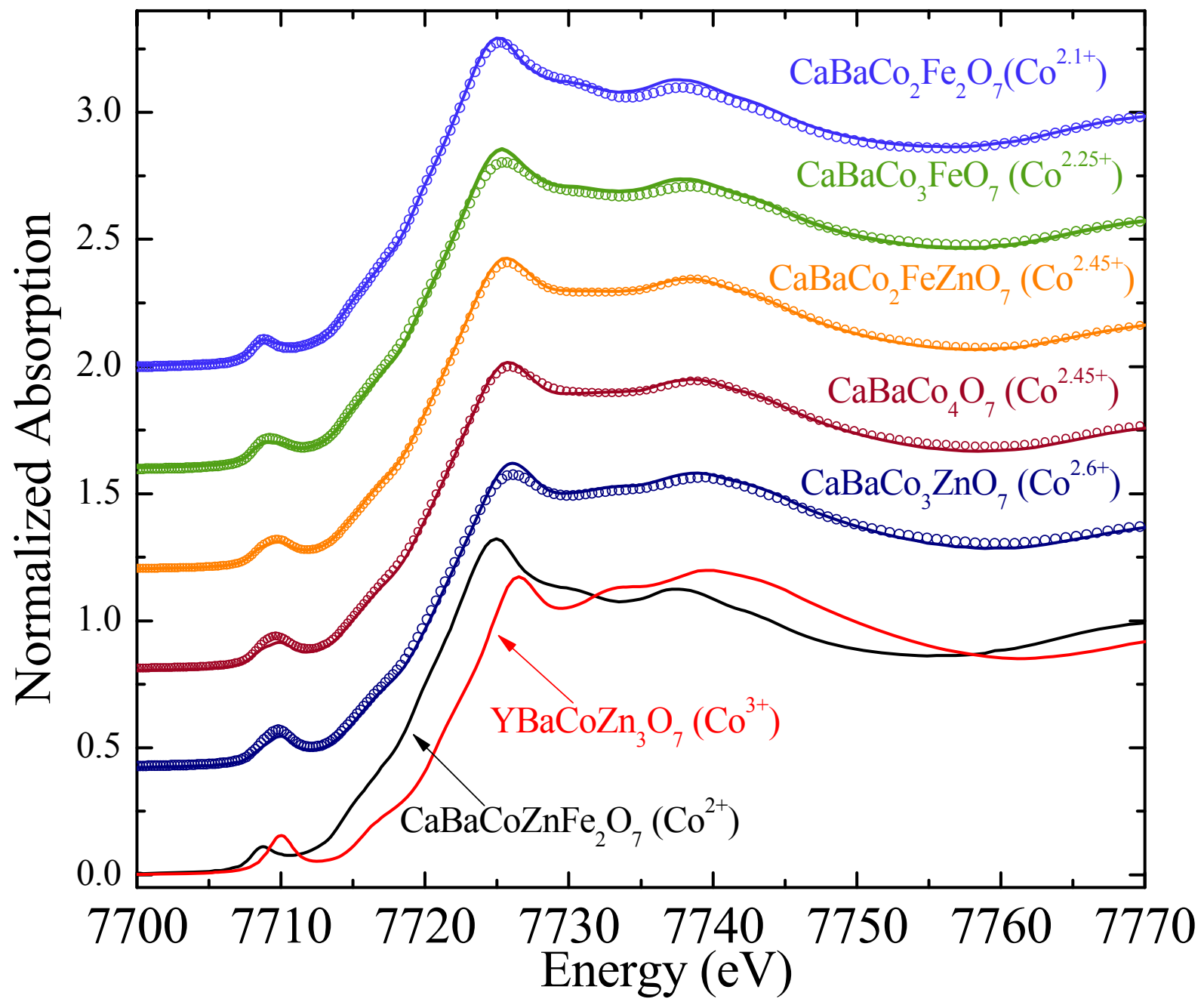

**Figure 10.** Comparison between the measured XANES spectra (points) and the best-fit weighted average sum (lines) of $Co^{3+}$ ($YBaCoZn_3O_7$) and $Co^{2+}$ ($CaBaCoZnFe_2O_7$) reference XANES spectra, plotted on the bottom part of the graph.

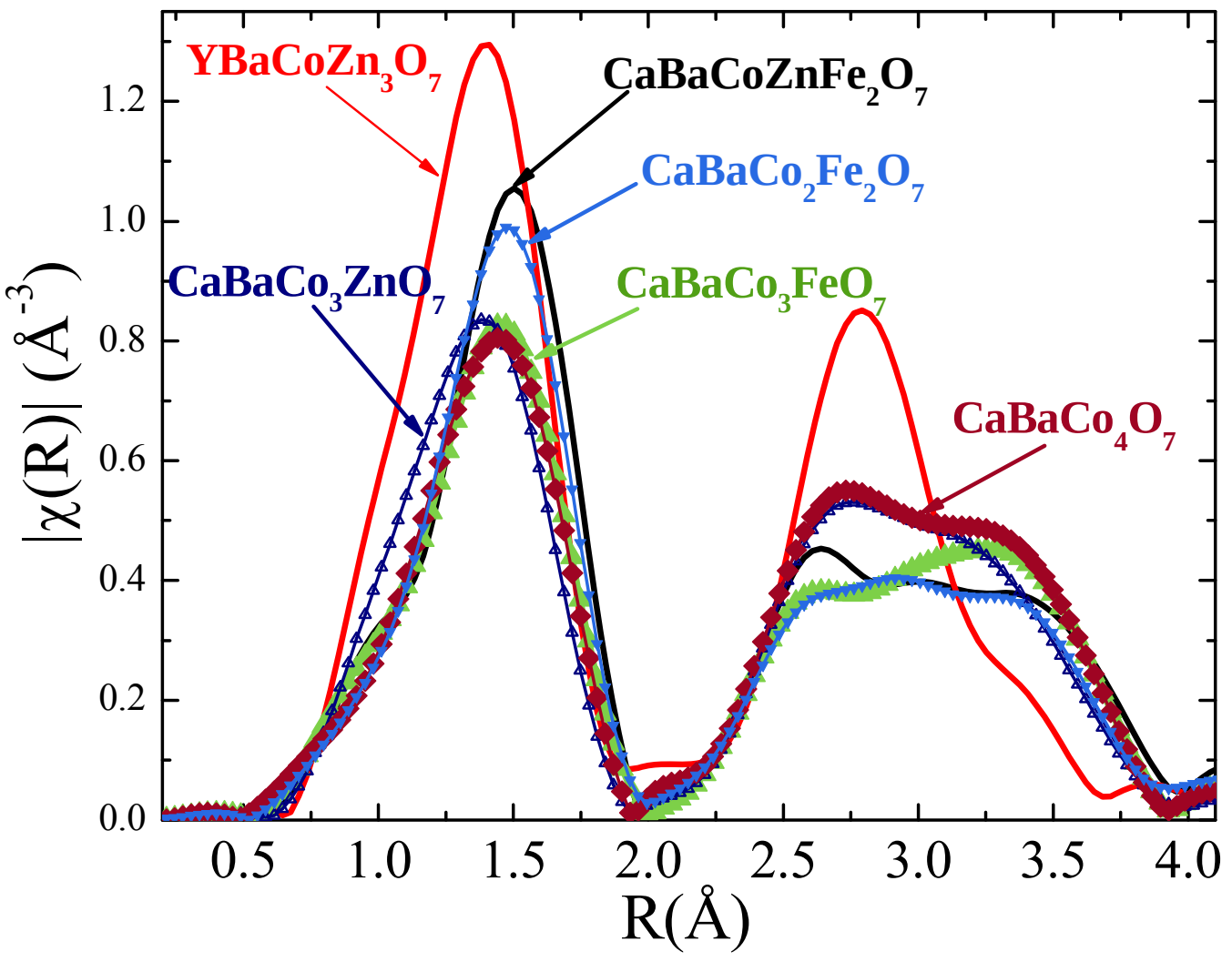

**Figure 11.** Modulus of the Fourier Transformed EXAFS signal weighted in $k^2$ (k: [1.8, 12.5], Δk=0.5) for $CaBaCoO_7$, $CaBaCo_{4-x}Fe_xO_7$ (x=1, 2) and $CaBaCo_3ZnO_7$ at T=35 K.

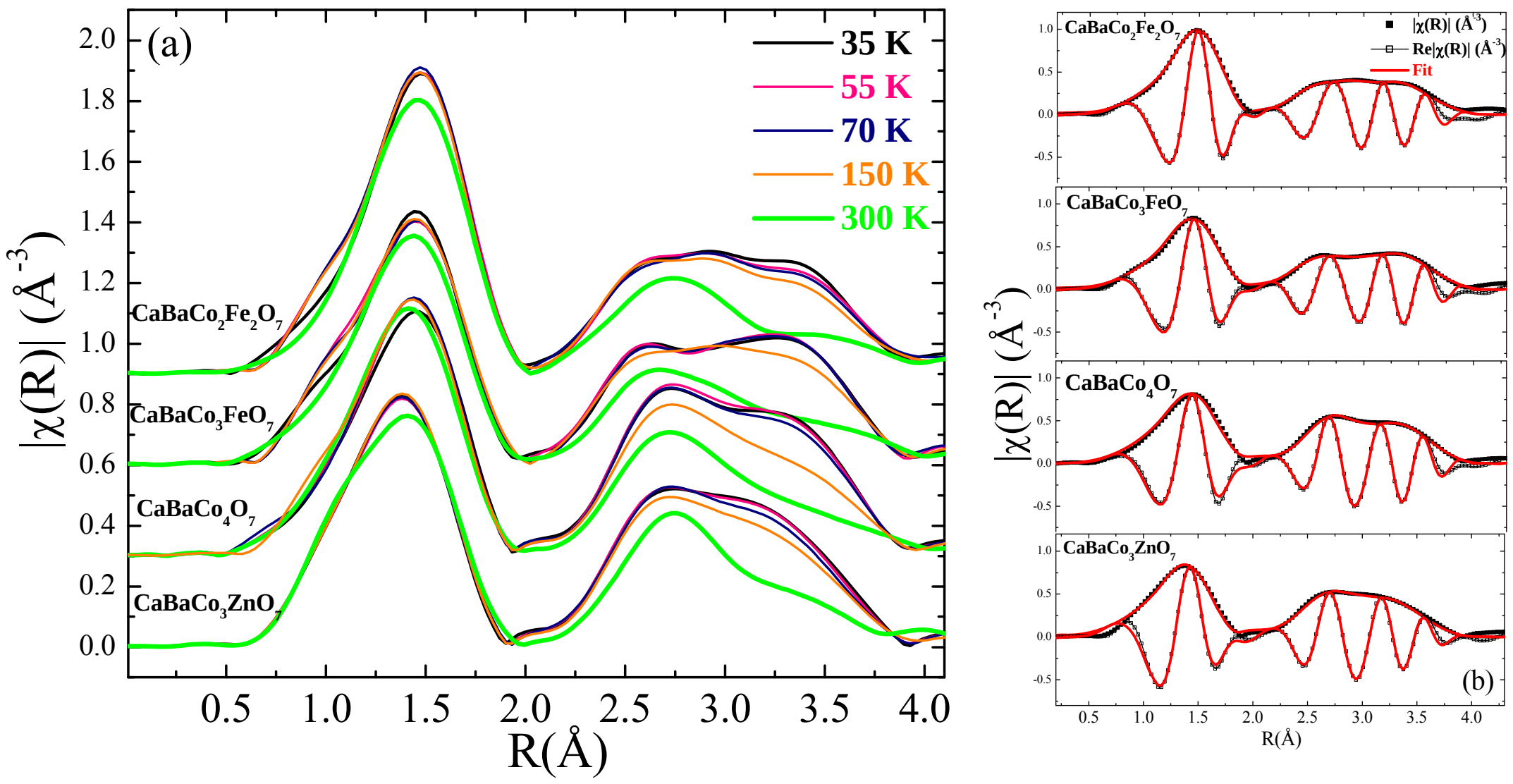

**Figure 12** (a) Modulus of the Fourier Transformed EXAFS signal weighted in $k^2$ (k: [1.8, 12.5], Δk=0.5) for $CaBaCoO_7$, $CaBaCo_{4-x}Fe_xO_7$ (x=1, 2) and $CaBaCo_3ZnO_7$ at different temperatures. The data of the first three samples have been shifted on the y-scale. (b) Modulus and real part of the Fourier Transformed EXAFS signal weighted in $k^2$ (k: [1.8, 12.5], Δk=0.5) and the corresponding fits (red lines) for $CaBaCoO_7$, $CaBaCo_{4-x}Fe_xO_7$ (x=1, 2) and $CaBaCo_3ZnO_7$ at T = 35 K.

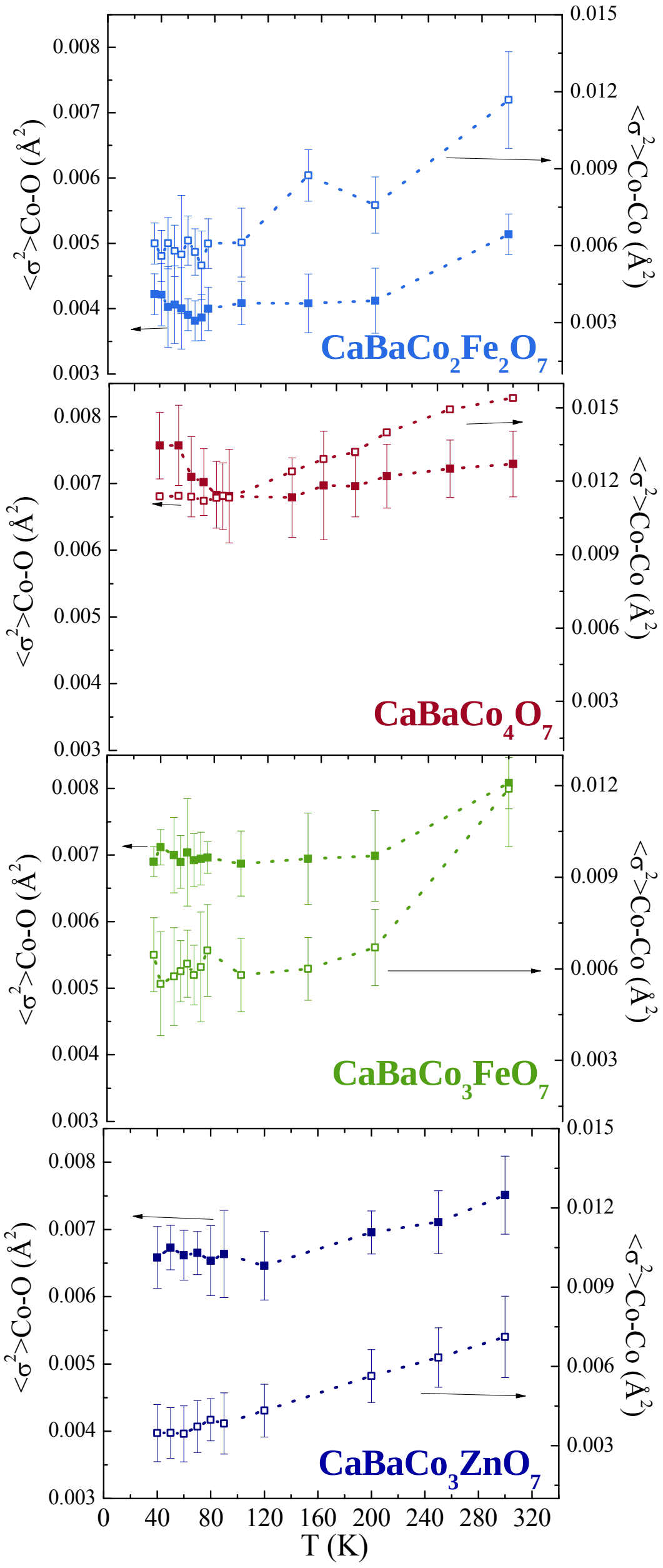

**Figure 13.** Temperature evolution of the Debye-Waller factors for $CaBaCo_4O_7$, $CaBaCo_{4-x}Fe_xO_7$ (x=1, 2) and $CaBaCo_3ZnO_7$ samples obtained from the fits up to R=4 Å. Filled squares indicate the Debye-Waller factors for the Co-O average first shell distance while open squares represent the Debye-Waller factors for the Co-Co average path length.

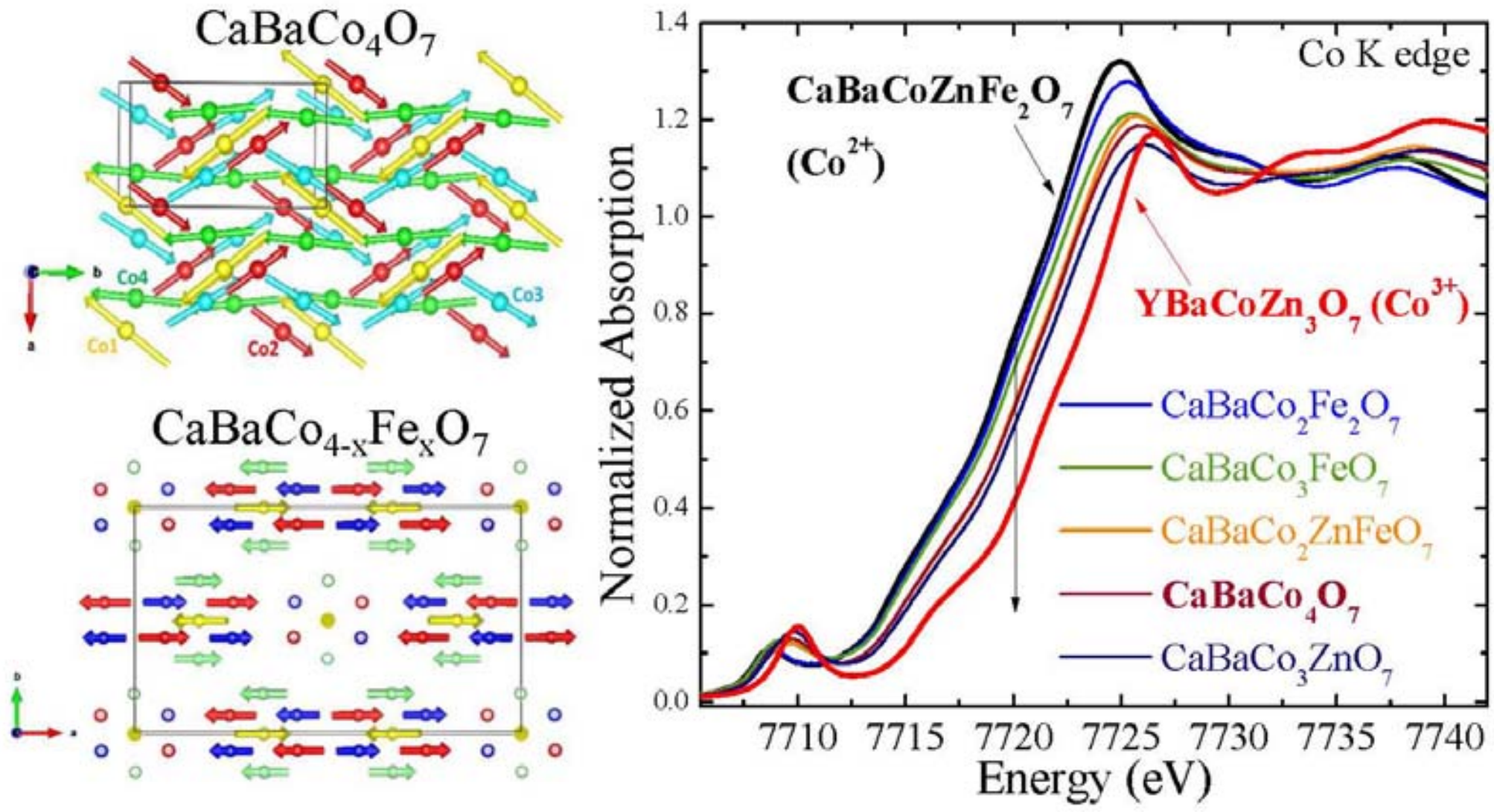

**For Table of Contents Only.** Left: magnetic structures of $CaBaCo_4O_7$ and $CaBaCo_{4-x}Fe_xO_7$ compounds. Right: X-ray absorption near-edge structure spectra from $CaBaCo_{4-x-y}Fe_xZn_yO_7$ and reference compounds.